\documentclass[prl,twocolumn,showpacs,byrevtex]{revtex4}
\input epsf
\usepackage{graphicx}
\usepackage{dcolumn}
\usepackage{bm}
\begin{document}
\preprint{}
\title{Radiative stability of neutrino-mass textures}
\author{M. K. Parida}
\email{mparida@nehu.ac.in}
\email{mparida@sancharnet.in}
\affiliation{Department of Physics, North Eastern Hill University, Shillong
793022, India} 
\author{C. R. Das}
\email{crdas@email.com}
\affiliation{Department of Physics, North Eastern Hill University, Shillong
793022, India}
\author{G. Rajasekaran}
\email{graj@imsc.ernet.in}
\affiliation{Institute of Mathematical Sciences, Chennai 600113, India}
\date{\today}
\begin{abstract}
Neutrino-mass textures proposed at high-scales are known to be unstable against
radiative corrections especially for nearly degenerate eigen values. Within the
renormalization group constraints we find a mechanism in a class of gauge 
theories which guarantees reproduction of any high-scale texture at low 
energies with radiative stability. We also show how the mechanism explains
solar and atmospheric neutrino anomalies through the bimaximal texture at high
scale.
\end{abstract}
\maketitle
\textbf{I.~Introduction:}
A major challenge to particle physics at present is the theoretical
understanding of experimentally observed neutrino anomalies. This has led
to suggestions of many interesting models and mass textures with hierarchial 
or degenerate eigen values \cite{1,2,3,4,5,6,7,8}. Whereas the observed 
mixing between quarks are small, experimental indications appear to favor 
maximal mixings in the neutrino sector. A possible mechanism to explain large 
neutrino mixings at low-energies starting from small high-scale mixings similar
to quarks could be through radiative magnification for quasidegenerate 
neutrinos \cite{9,10}.
\par 
An outstanding problem with bimaximal neutrino-mass textures with degenerate
eigen values is the instability of the masses and mixing angles due to 
radiative corrections which spoils their prospects for the 
neutrinoless-double-beta decay and the neutrino anomalies \cite{10,11,12}. 
While investigating radiative stability the usual procedure has been to assume
the bimaximal texture to be associated with a single 5-dim operator through 
see-saw mechanism with SM or MSSM as gauge theories at lower scales 
\cite{9,10,11,12}.
\par
Using renormalization group constraints in this letter we show that it is 
possible to reproduce any high-scale mass texture at low scale $(\mu=M_Z)$ 
with high degree of accuracy leading to stable evolution of the physically 
relevant Majorana-neutrino-mass matrix. The models where the mechanism 
operates consists of two component matrices contributing to the resultant
Majorana-neutrino mass texture at the highest scale. It is quite interesting 
to note that the mechanism operates successfully in 2HDM and also with SM and 
MSSM in the presence of type II see-saw mechanism probing left-right model 
(LRM) and $SO(10)$ GUT as prospective high-scale theories \cite{6,13,14}. We 
also show how experimental data on solar and atmospheric neutrino anomalies are
explained through the bimaximal texture.
\par
\textbf{II.~The mechanism in 2HDM:}
In the SM, MSSM, and 2HDM, the Majorana-neutrino mass in the flavor basis 
originates from 5-dim operators generated at the lepton-number-breaking scale
$(\mu_0\simeq M_N,\,t_0=\ln M_N)$ through see-saw mechanism. In the SM there 
is only one Higgs doublet $\Phi$ with VEV $\langle\Phi\rangle_0=v/\sqrt{2}$ 
and one operator $(=K^{(SM)})$ contributing to the Majorana mass 
$m^{(SM)}=-(1/4)K^{(SM)}v^2$. In the MSSM, although there are two Higgs 
doublets, there is only one neutrino-mass operator contributing to the 
neutrino-mass matrix due to the coupling of up type doublet with 
$m^{(MSSM)}=-(1/4)K^{(MSSM)}v_u^2$. In a class of 2HDM there are two doublets,
$\Phi_u$ and $\Phi_d$, with VEVs $v_u/\sqrt{2}=v\sin\beta/\sqrt{2}$, 
$v_d/\sqrt{2}=v\cos\beta/\sqrt{2}$. But, unlike SM or MSSM, there are two 
neutrino-mass operators, $K^I$ and $K^{II}$, and two matrices $m^I$ and 
$m^{II}$ which add up to generate the physically relevant 
Majorana-neutrino-mass matrix \cite{10},
$m=-(1/ 4)\left(K^Iv^2_u+K^{II}v^2_d\right)\equiv m^I+m^{II}$.
We use the renormalization scheme where the runnings of $v_u$, $v_d$, and
$\tan\beta$ are ignored. Then the relevant RGEs and their one-loop solutions 
for $\mu<M_N\,(t=\ln\mu<t_0)$ are,
\begin{widetext}
\begin{eqnarray}
16\pi^2{dm^I\over dt}&=&\left\{-3g_2^2+2\lambda_2+2{\rm Tr}\left(3Y_U^\dagger 
Y_U\right)\right\}m^I+{1\over 2}\left[\left(Y_E^\dagger Y_E\right)m^I+m^I
\left(Y_E^\dagger Y_E\right)^T\right]+2\lambda_3m^{II},\nonumber\\
16\pi^2{dm^{II}\over dt}&=&\left\{-3g_2^2+2\lambda_1+2{\rm Tr}\left(3
Y_D^\dagger Y_D+Y_E^\dagger Y_E\right)\right\}m^{II}-{3\over 2}\left[\left(
Y_E^\dagger Y_E\right)m^{II}+m^{II}\left(Y_E^\dagger Y_E\right)^T\right]+2
\lambda_3^*m^I,\nonumber\\
m_{ij}^I(t)&=&a_{ij}^I(t)m_{ij}^I(0),\;
a_{ij}^I(t)=I_{g_2}^{-{3\over 2}}I_{\lambda_2}I_{top}^3\left(I_iI_j\right)
^{1\over 4}R_{ij},\label{eq6}\\
m_{ij}^{II}(t)&=&a_{ij}^{II}(t)m_{ij}^{II}(0),\;
a_{ij}^{II}(t)=I_{g_2}^{-{3\over 2}}I_{\lambda_1}I_{b}^3I_\tau\left(I_iI_j
\right)^{-{3\over 4}}{\tilde R}_{ij}.\label{eq7}
\end{eqnarray}
\end{widetext}
Here $m_{ij}^{I,II}(0)=m_{ij}^{I,II}(t_0)$,
\begin{eqnarray}
I_l&=&\exp\left({1\over 8\pi^2}\int_{t_0}^th_l^2dt\right),
(l=e,\mu,\tau,top,b)\nonumber\\
I_{g_k}&=&\exp\left({1\over 8\pi^2}\int_{t_0}^tg_k^2dt
\right),(k=1,2)\nonumber\\
I_{\lambda_k}&=&\exp\left({1\over 8\pi^2}\int_{t_0}
^t\lambda_kdt\right),(k=1,2)\nonumber\\
R_{ij}&=&\exp\left[{1\over 8\pi^2}\int^t_{t_0}\left(m^{II}{m^I}^{-1}\right)_
{ij}\lambda_3dt\right],\nonumber\\
\tilde{R}_{ij}&=&\exp\left[{1\over 8\pi^2}\int^t_{t_0}\left(m^I{m^{II}}^{-1}
\right)_{ij}\lambda^*_3dt\right].\label{eq8}
\end{eqnarray}
Any texture at the highest scale for the physically relevant 
Majorana-neutrino-mass matrix
\begin{equation}
m(0)=m^I(0)+m^{II}(0),\label{eq9}
\end{equation}
never determines both the matrices $m^I(0)$ and $m^{II}(0)$. Given any element 
$m_{ij}(0)$ at $t_0$, one of the component elements, $m^I_{ij}(0)$ or 
$m^{II}_{ij}(0)$, remains completely undetermined at that scale. Then 
(\ref{eq6})-(\ref{eq7}) show that the same matrix, $m^I(t)$ or $m^{II}(t)$, is
undetermined at all lower scales too. This is in clear contrast to the cases 
in conventional analyses (CA) with SM or MSSM where there is only one $m(0)$ 
at $\mu=M_N$ and the texture gives all the elements of $m_{ij}(0)$ and 
$m_{ij}(t)$ \cite{10,11}. Now we impose the stability criterion that the 
texture is exactly reproduced at the lowest scale by demanding that
\begin{equation}
m_{ij}(t_Z)=m_{ij}^I(t_Z)+m_{ij}^{II}(t_Z)\equiv m_{ij}(0).
\label{eq10}
\end{equation}
Since $a^I(t_z)$ and $a^{II}(t_Z)$ are known in terms of the model parameters,
solutions of (\ref{eq9}) and (\ref{eq10}) now determine both $m^I(0)$ and 
$m^{II}(0)$ in terms of the high-scale neutrino-mass texture, $m(0)$,
\begin{eqnarray}
m^I_{ij}(0)&=&m_{ij}(0)\left(a^{II}_{ij}(t_Z)-1\right)/d_{ij},\nonumber\\
m^{II}_{ij}(0)&=&m_{ij}(0)\left(1-a^I_{ij}(t_Z)\right)/d_{ij},\nonumber\\
d_{ij}&=&a^{II}_{ij}(t_Z)-a^I_{ij}(t_Z).\label{eq11}
\end{eqnarray}
These parameters of the component matrices, determined from the boundary 
conditions (\ref{eq9}) and (\ref{eq10}) are expected to guarantee reproduction
of the high-scale texture at $M_Z$ when $m^I(t)$ and $m^{II}(t)$ are evolved 
through (\ref{eq6})-(\ref{eq8}).
\par
\begin{figure}
\epsfxsize=8.5cm
\epsfbox{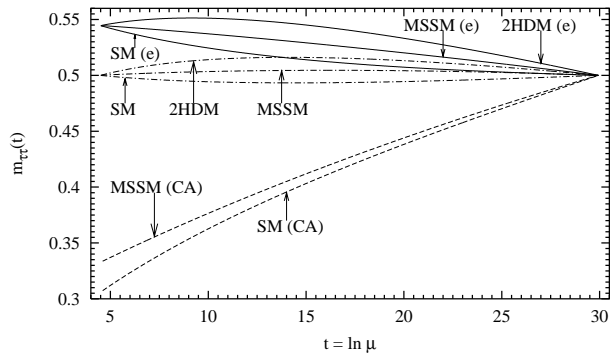}
\caption{RG-stable evolutions of $m_{\tau\tau}(t)$ in 2HDM, MSSM with 
$\tan\beta=40$, and SM. MSSM (CA) and SM (CA) denote conventional analyses in 
MSSM and SM having only one matrix at the highest scale. 2HDM (e), MSSM (e) and
SM (e) represent RG evolutions matching the experimental data on neutrino 
anomalies.}
\label{fig1}
\end{figure}

As an example we study RG evolution of the bimaximal texture with triply 
degenerate masses at $M_N\simeq 10^{13}$ GeV \cite{4}
\begin{equation}m(0)=\left[\begin{array}{ccc}
0&{1\over\sqrt{2}}&{1\over\sqrt{2}}\\
{1\over\sqrt{2}}&{1\over 2}&{-{1\over 2}}\\
{1\over\sqrt{2}}&{-{1\over 2}}&{1\over 2}\end{array}\right]m_0.\label{eq12}
\end{equation}
Using, $\lambda_1=0.16,\,\lambda_2=1.13,\,\lambda_3\simeq -0.011$ \cite{10}, 
and $\tan\beta=40$ we obtain $I_{\lambda_1}\simeq 0.95$, 
$I_{\lambda_2}\simeq 0.6976$, $I_{top}=0.833213$, $I_b=0.935023$, 
$I_\tau=0.950882$, $I_\mu=0.999832$, $I_e=0.999999$, $I_{g_2}=0.478614$ at 
$\mu=M_Z$. We compute $a^I(t_Z)$ and $a^{II}(t_Z)$, and, hence, $m^I(0)$ and 
$m^{II}(0)$ shown in Table \ref{tab1} as input parameters. The solutions for 
$m^I(t)$ and $m^{II}(t)$ are obtained through (\ref{eq6})-(\ref{eq8})and the 
elements of the Majorana-neutrino-mass matrix $m_{ij}(t)$ are obtained as 
their sum for all $t<t_0$. In Fig.~\ref{fig1} we have shown the radiative 
corrections for $m_{\tau\tau}(t)$. For comparison we have shown the results of
the conventional analysis as SM (CA) and MSSM (CA) for which there is only one 
matrix at the highest scale. The maximum radiative correction of the matrix 
elements in 2HDM using the present mechanism is found to be only 3-4\% as 
compared to 30-40\% in the SM (CA) or MSSM (CA). Whereas the maximal 
corrections in SM (CA) or MSSM (CA) occur at $\mu=M_Z$, in our case they occur 
with substantially reduced magnitude at intermediate scales. NonSUSY SM and 
2HDM have been successfully embedded in $SO(10)$ with single intermediate 
symmetries \cite{15}.
\par
\textbf{III.~Implementation in SM or MSSM:}
We note that the present mechanism also operates in SM and MSSM if they 
originate from high-scale theories which predict two component matrices at 
$M_N$. The popular see-saw mechanism which has its natural origin in LRM and 
$SO(10)$ contains the second contribution and leads to the two matrices in type
II see-saw formula with $m^I=m^{\rm SM}(m^{\rm MSSM})$ and 
$m^{II}\simeq fv^2/M_N(fv^2_u/M_N)$ when SM (MSSM) is obtained after symmetry 
breaking of LRM or $SO(10)$ \cite{1,6,7,8,14,15,16}. Here $f$ is the Majorana 
type Yukawa coupling of the neutrino. The mechanism also operates in SM
or MSSM when there are other types of contributions \cite{16}. The problem of 
obtaining a specific texture for $m(0)$, or $m^I(0)$ and $m^{II}(0)$, may call
for appending specific flavor symmetries to 
$SU(2)_L\times SU(2)_R\times U(1)_{B-L}\times SU(3)_C$ or $SO(10)$. Assuming 
such possibilities we derive the constraints on $m^I(0)$ amd $m^{II}(0)$ 
resulting from the bimaximal texture for $m(0)$ and its radiative stability. 
The RG evolutions of the standard see-saw term is the same as in SM or MSSM as
shown through $a^I(t)$ in (\ref{eq14})-(\ref{eq16}) below. But those for 
$m^{II}(t)$ occur due to loop-mediation of the standard-weak-Higgs doublet and
gauge bosons (plus superpartners) with the LH neutrinos alone. We derive them 
as
\begin{eqnarray}
16\pi^2{dm^{II}\over dt}&=&\left(c^{(1)}g_1^2+c^{(2)}g_2^2\right)m^{II}
\nonumber\\
&&+c^{(3)}\left[\left(Y_E^\dagger Y_E\right)m^{II}+
m^{II}\left(Y_E^\dagger Y_E\right)^T\right],\nonumber
\end{eqnarray}
where $c^{(i)}=(9/10,3/2,-3/2)$ for SM, but $c^{(i)}=(-9/5,-9/2,1)$ for MSSM. 
In the notations of (\ref{eq6})-(\ref{eq7}) we obtain\\
\underline{SM}
\begin{eqnarray}
a_{ij}^I(t)&=&I_{g_2}^{-{3\over 2}}I_\lambda I_{top}^3I_b^3I_\tau
\left(I_iI_j\right)^{-{3\over 4}},\nonumber\\
a_{ij}^{II}(t)&=&I_{g_1}^{9\over 20}I_{g_2}^{3\over 4}
\left(I_iI_j\right)^{-{3\over 4}},\label{eq14}
\end{eqnarray}
\underline{MSSM}
\begin{eqnarray}
a_{ij}^I(t)&=&I_{g_1}^{-{3\over 5}}I_{g_2}^{-3}I_{top}^3
\left(I_iI_j\right)^{1\over 2},\nonumber\\
a_{ij}^{II}(t)&=&I_{g_1}^{-{9\over 10}}I_{g_2}^{-{9\over 4}}
\left(I_iI_j\right)^{1\over 2}.\label{eq16}
\end{eqnarray}
Then using (\ref{eq9})-(\ref{eq11}) we obtain the initial values of 
$m^{I,II}(0)$ and, hence, solutions for $m_{ij}(t)$ for $\mu=M_Z$ --- $M_N$ 
exhibiting stability of all the matrix elements of $m(t)$ under radiative 
corrections. The elements of the component matrices for the two cases are also
shown in Table \ref{tab1}. In Fig.~\ref{fig1} we have plotted 
$m_{\tau\tau}(t)$ in comparison to conventional analyses. As against the 
maximal 30-40\% radiative corrections in SM (CA) and MSSM (CA) occurring at 
$\mu=M_Z$, they are only 3-4\% in SM and 1.5\% in MSSM which occur at 
intermediate scales in the present analysis. Among all the three models, the 
minimum radiative corrections upto 1.5\% is found to occur in MSSM.
\par
\textbf{IV.~Fitting the neutrino anomalies:}
When the bimaximal texture is exactly reproduced at $M_Z$, one way to explain
neutrino anomalies could be through threshold effects \cite{17}. But here 
ignoring threshold effects we show how the present mechanism permits matching 
of the observed solar (LAMSW) and atmospheric neutrino anomalies starting from
the bimaximal texture with degenerate mass eigen values at the highest scale. 
Using quasidegenerate neutrinos with masses $m_1=-0.2$ eV, $m_2=0.200045$ eV, 
$m_3=0.2075$ eV which are spread around $m_0=0.2$ eV, the mixing angles 
suitable for LAMSW with $s_3=0.6946$ and atmospheric neutrino oscillations with
$s_1=0.6950$, it is straight forward to construct the mass matrix consistent 
with the experimental data
\begin{equation}
m^{(e)}(t_Z)=\left[\begin{array}{rrr}
-0.044716&0.722025&-0.690426\\  
0.722025&0.501718& 0.477908\\  
-0.690426&0.477908& 0.544506
\end{array}\right]m_0.\label{eq17}
\end{equation}
Although we have used $s_2=0$, the mechanism is found to work for other values
consistent with CHOOZ bound \cite{18}. Similarly the mechanism also works with
other values of $m_0\simeq 0.1-1.0$ eV. Within the RG-constraints, the high 
scale texture can match the experimentally observed anomalies provided 
$m_{ij}(0)$ in (\ref{eq10}) and (\ref{eq11}) is replaced by $m_{ij}^{(e)}(t_Z)$
leading to
\begin{eqnarray}
m^I_{ij}(0)&=&\left(a^{II}_{ij}(t_Z)m_{ij}(0)-m_{ij}^{(e)}(t_Z)\right)/d_{ij},
\nonumber\\
m^{II}_{ij}(0)&=&\left(m_{ij}^{(e)}(t_Z)-a^I_{ij}(t_Z)m_{ij}(0)\right)/d_{ij}.
\label{eq18}
\end{eqnarray} 
In Fig.~\ref{fig1} the curves 2HDM(e), MSSM(e) and SM(e) represent the result
of fitting the data through the high scale bimaximal texture given in
(\ref{eq12}) and $m^{(e)}(t_Z)$ given in (\ref{eq17}) using 2HDM, MSSM and SM,
respectively. We note that similar RG-stability also holds approximately for 
certain other elements depending upon the exact values of $s_1$ and $s_3$. But
the radiative corrections are found to be larger if the difference between 
$s_1$ and $s_3$ is larger. Similar curves can be plotted for other elements 
also.
\par
\textbf{V.~Conclusion:}
The present mechanism demonstrates how to evade RG-constraints on neutrino-mass
textures in conventional analyses. It operates in a class of gauge theories 
leading to 2HDM, SM or MSSM where two component matrices contribute to the 
physically relevant Majorana-neutrino mass at the highest scale. Once a 
resultant texture is generated using suitable flavor symmetries at the highest
scale, this mechanism determines the two unknown matrices at the highest scale
which ensure its RG-stability at all lower scales or its matching with the
experimental data. The mechanism can be applied to reproduce any high-scale 
texture at low energies with any desired degree of stability including higher 
order corrections in (\ref{eq10}). It is quite interesting that the stability 
criteria operate in the presence of type II see-saw mechanism and probe into 
models including left-right gauge theories and $SO(10)$ as prospective 
high-scale theories. The textures for component matrices derived from the
stability condition sets considerable constraint on future model building with
flavor symmetry.
\begin{table*}
\caption{Component matrices determined from RG stability criteria and also by
matching the experimental data as denoted by (e). Here $m_0$ is a common 
factor.}
\begin{ruledtabular}
\begin{tabular}{lcc}
Model&$m^I(0)/m_0$&$m^{II}(0)/m_0$\\\hline
2HDM&$\left[\begin{array}{rrr}
 0.000000&-0.220454&-0.146400\\ 
-0.220454&-0.155695& 0.103358\\ 
-0.146400& 0.103358&-0.058182
\end{array}\right]$
&$\left[\begin{array}{rrr}
0.000000& 0.927561& 0.853507\\  
0.927561& 0.655695&-0.603358\\  
0.853507&-0.603358& 0.558182
\end{array}\right]$\\
&&\\
2HDM (e)&$\left[\begin{array}{rrr}
 0.112140&-0.257854& 3.041252\\ 
-0.257854&-0.160003&-2.126483\\ 
 3.041252&-2.126483&-0.151038
\end{array}\right]$
&$\left[\begin{array}{rrr}
-0.112140&0.964961&-2.334145\\  
 0.964961&0.660003& 1.626483\\  
-2.334145&1.626483& 0.651038
\end{array}\right]$\\
&&\\
MSSM&$\left[\begin{array}{rrr}
0.000000& 0.414613& 0.389877\\  
0.414613& 0.293120&-0.275628\\  
0.389877&-0.275628& 0.257691
\end{array}\right]$
&$\left[\begin{array}{rrr}
0.000000& 0.292493& 0.317228\\  
0.292493& 0.206879&-0.224371\\  
0.317228&-0.224371& 0.242308
\end{array}\right]$\\
&&\\
MSSM (e)&$\left[\begin{array}{rrr}
0.061597& 0.394061& 2.364023\\  
0.394061& 0.290752&-1.657126\\  
2.364023&-1.657126& 0.193220
\end{array}\right]$
&$\left[\begin{array}{rrr}
-0.061597&0.313044&-1.656916\\  
 0.313044&0.209247& 1.157126\\  
-1.656916&1.157126& 0.306779
\end{array}\right]$\\
&&\\
SM&$\left[\begin{array}{rrr}
 0.000000&-0.292357&-0.292451\\ 
-0.292357&-0.206728& 0.206794\\ 
-0.292451& 0.206794&-0.206860
\end{array}\right]$
&$\left[\begin{array}{rrr}
0.000000& 0.999463& 0.999557\\  
0.999463& 0.706728&-0.706794\\  
0.999557&-0.706794& 0.706860
\end{array}\right]$\\
&&\\
SM (e)&$\left[\begin{array}{rrr}
 0.162944&-0.346719& 4.800250\\ 
-0.346719&-0.212991&-3.356768\\ 
 4.800250&-3.356768&-0.369052
\end{array}\right]$
&$\left[\begin{array}{rrr}
-0.162944&1.053825&-4.093143\\  
 1.053825&0.712991& 2.856768\\  
-4.093143&2.856768& 0.869052
\end{array}\right]$
\end{tabular}
\end{ruledtabular}
\label{tab1}
\end{table*}
\begin{acknowledgments}
The work of M.K.P. is supported by Project No.~SP/S2/K-30/98 and the work of
C.R.D. is supported by Project No.~98/37/9/BRNS-Cell/731 of the Government of
India.
\end{acknowledgments}

\end{document}